\documentclass[10pt]{article}
\usepackage{balance}
\usepackage{amsmath}
\usepackage{multicol}
\usepackage{graphicx}
\makeatletter
\newcommand\figcaption{\def\@captype{figure}\caption}
\makeatother
\usepackage{cite}
\usepackage{bm}
\usepackage{amsfonts}
\usepackage{indentfirst}
\newcommand{\ct}[1]{{\textsuperscript{{\cite{#1}}}}}

\newcommand{\bee}{\begin{equation}}
\newcommand{\ee}{\end{equation}}
\newcommand{\beea}{\begin{eqnarray}}
\newcommand{\eea}{\end{eqnarray}}
\newcommand{\fna}[1]{$\mathrm{\grave{#1}}$}
\newcommand{\fpi}[1]{$\mathrm{\acute{#1}}$}

\begin{document}
%\begin{CJK*}{GBK}{}
\title{A covering theory of special relativity}
\author{{\small  Daqing Liu$^1$\thanks{E-mail:liudq@mail.xjtu.edu.cn}, Xinghua Li$^2$ and Yanshen Wang$^1$}\\
  {\small  $^1$ Department of Applied Physics, Non-equilibrium Condensed
Matter and Quantum
        Engineering Laboratory, } \\
  {\small   KLME,   Xi'an Jiaotong University, Xi'an, 710049,
        China}\\
 {\small $^2$ Physics
 and Optic-Electronics Technology College, Fujian Normal University, Fuzhou 350007,
 China}
 %\ead{lixh@fjnu.edu.cn}
}
\date{}
\maketitle
%\end{CJK*}
% \begin{abstract}
\noindent{{\bf Abstract:} Under the assumption of closed-path
velocity of light invariant, we show both the general expression of
velocity of light in an ordinary inertial reference frame and the
generalized Lorentz transformation between the ordinary inertial
reference frame and the absolute (privileged) reference frame.
Although such assumption can not determine theory ambiguously, some
significant results can still be obtained by the assumption.
Furthermore, the study shows that the relativity of simultaneity is
not a universal concept. }

 \vskip 0.1in \noindent {{\bf
R$\mathbf{\acute{e}}$sum$\mathbf{\acute{e}}$:}
 Dans l'hypoth\fna{e}se se de la vitesse ferm\fpi{e} invariant chemin de
la lumi\fna{e}re, nous montrer \fna{a} la fois l'expression
g\fpi{e}n\fpi{e}rale de la vitesse de la lumi\fna{e}re dans un
cadre de
r\fpi{e}f\fpi{e}rence inertiel ordinaire et le transformation de
Lorentz g\fpi{e}n\fpi{e}ralis\fpi{e}e entre le cadre de
r\fpi{e}f\fpi{e}rence inertiel ordinaire et le cadre de
r\fpi{e}f\fpi{e}rence absolu. Bien que cette hypoth\fna{e}se ne
puisse pas d\fpi{e}terminer uniquement la th\fpi{e}orie, elle pourra
nous conduire \fna{a} certains r\fpi{e}sultats importants. En plus,
cette \fpi{e}tude montre que la relativit\fpi{e} de la
simultan\fpi{e}it\fpi{e} n'est pas un concept universel. }

%\noindent {\bf pacs:} 03.30.+p
%\end{abstract}
%\submitto {liudq@mail.xjtu.edu.cn}

\vskip 0.2in
\begin{multicols}{2}

\section{Introduction}
The special theory of
relativity\ct{einstein,stachel,holton,franklin,bondi,bergman}, one
of the most significant achievements of physics, is the cornerstone
of modern physics. It is the special theory of relativity that first
took space and time as a whole and introduced abstract reasoning
into the study of space-time. It has therefore improved our
understanding of the nature of space-time. In particular, compared
to the classical theory, there are many unique concepts in the
special relativity, such as Lorentz contraction, the relativity of
simultaneity and that there is no superluminal signal (Whether there
is superluminal signal is a hotspot topic of theoretical and
experimental studies\ct{5,6,7,8,9}). Essentially, all these unique
concepts stem from the two
postulates\ct{einstein,stachel,holton,franklin,bondi,bergman}:
relativity of inertial reference frames (relativity of IRFs), that
is, the physical laws are independent of the state of motion of the
reference frame, at least if the frame is not accelerated\ct{rel},
and the invariant of one-way velocity of light (one-way VL), that
is, the clocks can be synchronized in such a way that the
propagation velocity of light ray in vacuum, measured by means of
these clocks, becomes equal to a universal constant everywhere,
provided that the coordinate system is not accelerated\ct{vel}.

Bondi\ct{bondi}, Bergman\ct{bergman}, {\it etc.} claimed long time
ago that there are mismatches between the postulate of relativity
and cosmological observations. Cosmological observations tell us
that there does exist a distinguished IRF, to which many cosmology
phenomena respect\ct{levy}. Furthermore, as well known, what we have
measured is the two-way VL and the one-way VL depends on the
synchronization of the clocks. Different synchronization scheme
leads to different one-way VL and different form of theory even
two-way VL is invariant.

It is, then, interesting to study a more general theory, nominated
as covering theory\ct{covering,sexl}, than special relativity. The
covering theory, a parametric extension of special theory that
contradicts special relativity for all but one value of parameters,
is very useful for exploring logical implications and deciding on
the tradeoff between simplicity and predictive accuracy. There are
studies on the covering theory\ct{sexl,selleri,russo,serr2} with
two-way VL invariant\cite{russo,serr2} or without two-way VL
invariant\ct{sexl,selleri}.

These covering theories\ct{sexl,selleri,russo,serr2} are based on
the two points: 1) "general Lorentz transformation" or Robertson
transformation\ct{vargas} between ordinary inertial reference frame
and the absolute inertial reference; 2) the introduction of
synchronization parameter. The physical meaning of synchronization
parameter is obvious, of course. However, as shown in the refs.
\cite{sexl,selleri,russo}, such approach is always associated with a
lengthy deduction.

Here we want to consider the covering theory in a different
approach. In particular, we do not set the synchronization
parameter. We first generalize the
 VL invariant of backwards-and-forwards way into that of any closed
path. We assume that \emph{the closed-path VL is a universal
constant, independent on IRF and closed path}. Such observable
assumption can be reexpressed as that the closed-path light-travel
time is invariant under any shape, provided the length of the closed
path is fixed. The assumption, as shown in context, puts a strong
constraint on the covering theory. In fact, our study shows that
parameters in the constrained theory are only one more than that in
special relativity. We also investigate the validity of some unique
concepts in special relativity.

The manuscript is organized as follows. In section 2, we study the
velocity of light under the assumption. Then, the transformation
between ordinary IRF and absolute reference of frame (ARF) is
shown in section 3. The results are summarized in the last
section.

\section{Velocity of light}

As is known to all, one can not measure one-way VL unambiguously.
The logical situation becomes circular when one tries to measure
one-way VL: On one hand to measure one-way VL one needs to synchronize
clocks, while on the other hand to synchronize clocks one must know
the one-way VL. To avoid the synchronization problem among clocks
one may, for instance, use one clock.

Thus, all the laboratory experiments measured instead the
two-way VL. Consider the following situation: In an inertial frame
$\Sigma$, a flash of light leaves point A at time $t_1$, is
reflected back in point B at times $t_2$, and returns point $A$ at
time $t_3$. Suppose
VL along vector $\stackrel{\rightarrow}{AB}$ is
$c(\stackrel{\rightarrow}{AB})$, VL along vector
$\stackrel{\rightarrow}{BA}$ is $c(\stackrel{\rightarrow}{BA})$ respectively, we
have
 \bee
 c(\stackrel{\rightarrow}{AB})=\frac{|\stackrel{\rightarrow}{AB}|}{t_2-t_1},\, \,
 c(\stackrel{\rightarrow}{BA})=\frac{|\stackrel{\rightarrow}{BA}|}{t_3-t_2},
 \ee
 where
$|\stackrel{\rightarrow}{AB}|=|\stackrel{\rightarrow}{BA}|$ and  VLs, $c(\stackrel{\rightarrow}{AB})$ and $c(\stackrel{\rightarrow}{BA})$,
 depend on the propagation directions and IRF .
In the above equation to determine $t_2-t_1$ and $t_3-t_2$ one needs to
synchronize the clock at point A and clock at point B. However, if one
considers only $t_3-t_1$ the synchronization is not needed. Therefore, since $t_3-t_1=t_3-t_2+t_2-t_1$£¬ one can define
two-way VL as
 \bee \label{two-way}
 \bar{c}=\frac{|\stackrel{\rightarrow}{AB}|+|\stackrel{\rightarrow}{BA}|} {t_3-t_1}
 =\frac{|\stackrel{\rightarrow}{AB}|+|\stackrel{\rightarrow}{BA}|} {\frac{|\stackrel{\rightarrow}{AB}|}{c(\stackrel{\rightarrow}{AB})}
 +\frac{|\stackrel{\rightarrow}{BA}|}{c(\stackrel{\rightarrow}{BA})}}
 \ee
 where $|\stackrel{\rightarrow}{AB}|+|\stackrel{\rightarrow}{BA}|$ is the total length of the back-and-forward path, a special case of closed path, and $\frac{|\stackrel{\rightarrow}{AB}|}{c(\stackrel{\rightarrow}{AB})}
 +\frac{|\stackrel{\rightarrow}{BA}|}{c(\stackrel{\rightarrow}{BA})}$ is the total light-travel time along the back-and-forward path. Such measurement is just with one clock and irrespective
 of the synchronization. So far all the experiments point out that
 $\bar{c}=3\times 10^8m/s$ is a universal constant, that is, it is dependent neither on the direction
 and distance of $\stackrel{\rightarrow}{AB}$ nor on the IRF.

Such definition of two-way VL can be generalized to that of
arbitrary closed-path VL. Suppose in a closed-path we choose a
infinitesimal arc with the length $dl$, the total length of the
closed path is $\oint dl$. We then analogously define the close-path
VL as
 \bee \label{define}
 \bar{c}=\frac{\oint dl}{t},
 \ee
 where $t$ is the total light-travel time along the closed path. It is easy to see that
 Eq. (\ref{define}) is a direct generalization of Eq. (\ref{two-way}) and
such definition of the closed-path VL $\bar{c}$ is irrespective of
the synchronization. We think such generalization is very natural and two-way VL is a special case of closed-path VL.
 Suppose again that VL along the infinitesimal arc is $c(\varphi)$,
 where $\varphi$ is the propagation direction of the light, $t$ can
be expressed as $t=\oint \frac{dl}{c(\varphi)}$. One finds that
 \bee
\tag{\ref{define}$'$}
 \label{define:prime}
 \bar{c}=\frac{\oint dl}{\oint \frac{dl}{c(\varphi)}}.
 \ee

Since back-and-forward path is a special closed path and such
two-way VL is a universal constant $3\times 10^8 m/s$, we assume in
this paper that closed-path velocity of light is also the same
constant, which is independent on IRF and closed path.
%For simplification we set $\bar{c}=1$.
Thus from Eq. (\ref{define:prime}), with $\bar{c}$ universal
constant, we have
 \bee \label{relation}
\oint \frac{dl}{c(\varphi)}= \frac{1}{\bar{c}}{\oint dl}
 \ee
 for arbitrary closed path in any IRF.

For simplification we first choose the closed path as a smooth plane
convex curve, for instance, in x-y plane. Then, VL should be a
periodic function of the angle, $\varphi\in [0,2\pi]$, between
propagation direction and a fixed direction, in particular, x-axis.
As shown in Fig.1, we consider an infinitesimal arc,
$\stackrel{\frown}{AB}$, the length of which is
$dl=|\stackrel{\frown}{AB}|$, on the closed curve. Suppose the
radius of the arc curvature is $R$ and the arc angle expanded to the
corresponding curvature center is $d\theta$, one has $dl=Rd\theta$.
We assume the angle between the tangent of curve at point A and
x-axis (or polar axis) is $\varphi$ and the angle between the
tangent at point B and x-axis is $\varphi+d\varphi$ respectively. It
is easy to see that $d\varphi=d\theta$. If the velocity of light
along the arc is $c(\varphi)$, the travel of light along the
infinitesimal arc needs time interval
 \bee
 dt=\frac{Rd\theta}{c(\varphi)}=\frac{Rd\varphi}{c(\varphi)}.
 \ee

\begin{center}
~\\[\intextsep]
\begin{minipage}{0.46\textwidth}
\centering
%\vskip 0.2in
\includegraphics[width=1.8in]{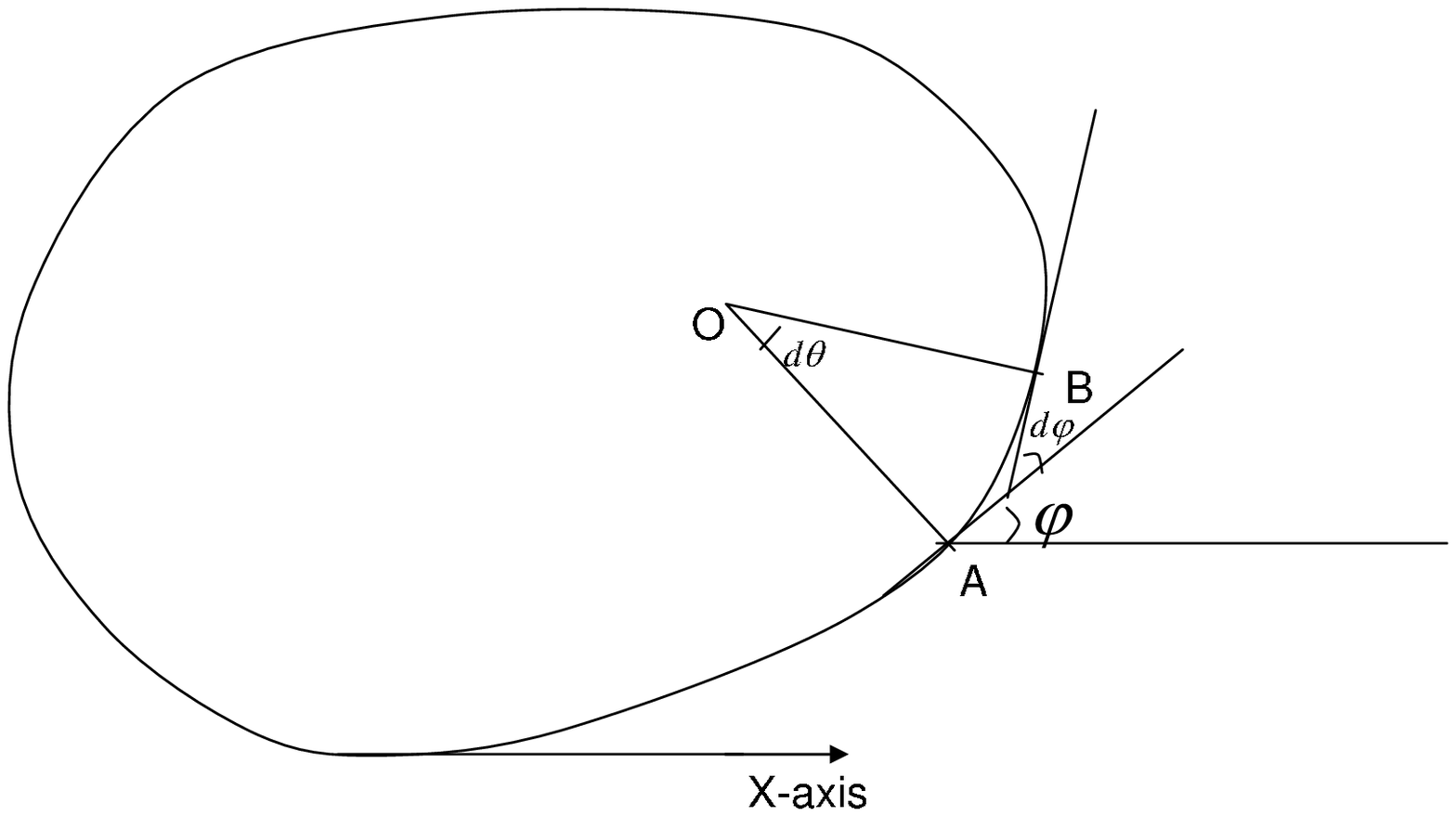}%
\vskip 0.1in \figcaption{Travel of light along a closed path. VL is a function of angle $\varphi$.} \label{lattice}
\end{minipage}
\setlength{\intextsep}{0.3in plus 0.1in minus 0.2in}
\\[\intextsep]
\end{center}

Since the closed curve is convex, $\varphi$ and point belonging to
the closed curve are one-to-one. From Eq. (\ref{relation}) we get
 \bee \label{loop}
 \oint \frac{R(\varphi)d\varphi}{c(\varphi)} =\frac{1}{\bar{c}} \oint
 R(\varphi)d\varphi,
 \ee
where observable quantity $\bar{c}$ is a university constant. This
equation is not an identical one but a direct deduction of equation
(\ref{relation}).

In this paper we assume the universal constant $\bar{c}\equiv 1$ for
simplification, we have then
 \bee
\tag{\ref{loop}$'$}
 \label{loop:prime}
 \oint R(\varphi) f(\varphi)d\varphi=\int_0^{2\pi}R(\varphi) f(\varphi)d\varphi \equiv 0,
 \ee
where $f(\varphi)=\frac{1}{c(\varphi)}-1$ is a function of $\varphi$
with period $2\pi$. For a closed convex curve in $x-y$
plane, radius of curvature $R(\varphi)$
 is not an arbitrary positive function. It should and should only meet the requirements,
 \beea
 \oint dx &=&\int_0^{2\pi}R(\varphi) \cos\varphi d\varphi =0,
 \nonumber \\
 \oint dy &=&\int_0^{2\pi}R(\varphi) \sin\varphi d\varphi =0.
 \label{curvature}\eea
At the same time, periodic function $f(\varphi)$ can be expanded by Fourier series,
 \bee \label{function_f}
 f(\varphi)=a_0+ \sum\limits_{n=1}^\infty
 (a_n\cos n\varphi+b_n\sin n\varphi)
 \ee
We emphasize again that Eq. ({\ref{loop:prime}}) puts a very strong
constraint on the VL. Substituting Eqs. (\ref{curvature}) and
(\ref{function_f}) into expression ({\ref{loop:prime}}), one finds
 \bee
 \int\limits_0^{2\pi}d\varphi R(\varphi)[a_0+ \sum\limits_{n=2}^\infty
 (a_n\cos n\varphi+b_n\sin n\varphi)]=0
 \ee
for arbitrary positive function $R(\varphi)$ satisfying
(\ref{curvature}). One can prove that to meet the above requirement
all $a_0$, $a_n$, $b_n$ ($n=2,\,3,\cdots\infty$) should be zero. For
instance, to prove $a_0=0$ one may choose $R = 1$. Therefore,
$f(\varphi)$ has only two free parameters,
$f(\varphi)=A_1\cos\varphi+B_1\sin\varphi\doteq A_0
\cos(\varphi+\varphi_0)$, where $A_1$ and $B_1$ (or $A_0$ and
$\varphi_0$) are constant depending on the plane and IRF. It is easy
to see that as long as $A_0$ is fixed, we can get all the VL
provided the propagation direction is in the plane. In particular,
in the case of $A_0=0$, VL is independent on the propagation
direction, that is, it keeps invariant in the propagation plane. For
$A_0\neq 0$, we are able to choose a suitable polar axis to make
$\varphi_0=0$. We have, then, three obvious conclusions: 1) VL is
maximum(minimum) along polar axis if $-1<A_0<0$($0<A_0<1$). 2) The
direction of the maximum VL and that of the minimum VL are
anti-parallel. 3) The direction of the minimum (maximum) VL is
unique in the plane.

For an inertial reference frame, VL is constant unity provided
$A_0=0$ for all the plane. Such isotropic inertial reference is
nominated as absolute reference frame(ARF). Meanwhile, for an
ordinary inertial reference frame in which VL varies with the
propagation direction, we can always find a direction along which VL
is minimum. This direction is also unique, otherwise one would
obtain a contradictory result against conclusion 3) in the above
paragraph.

If we choose such direction as our polar axis of the IRF,
VL can be written as
 \bee \label{eq5}
c(\varphi)=\frac{1}{1+A\cos\varphi},
 \ee
where $A>0$ is a constant which depending on the inertial reference
frame and $\varphi$ is the angle between the propagation direction
and the polar axis. It is not difficult to prove that the
closed-path VL is indeed constant unity for arbitrary closed path,
including spatial closed path, provided VL is expressed by the above
formulae.

Here we show a simple proof in cartesian coordinate. We set polar
axis coincides with  $z$-axis, that is, $\varphi$ is the angle
between the propagation direction and $z$-axis.  Without loss of
generality, we assume that the closed path is a smooth curve, and
choose a infinitesimal smooth arc along the propagation direction on
the curve, $\stackrel{\frown}{AB}$, where the coordinates of $A$ and
$B$ are $(x,y,z)$, $(x+dx,y+dy,z+dz)$ respectively. The
infinitesimal arc can be approximated by an infinitesimal vector
$\stackrel{\rightarrow}{AB}$. Thus the propagation direction of the
light is $\stackrel{\rightarrow}{AB}=(dx,dy,dz)$. Suppose the
length of $\stackrel{\rightarrow}{AB}$ is $dl$, one finds that $\cos
\varphi=\frac{dz}{dl}$. Therefore, the travel of light along the
infinitesimal arc needs time interval \bee
dt=\frac{dl}{c(\varphi)}=(1+A\cos \varphi)dl=dl+Adz, \ee where we
have used the Eq. (\ref{eq5}) and the expression of $\cos\varphi$.
Thus, travel of light along the closed path needs time \bee t=\oint
dt=\oint (dl+Adz)=\oint dl, \ee provided the curve is closed, that
is, $\oint dz=0$.

\section{Transformation between ARF and IRF}
Under the assumption of closed-path VL invariant we deduce in the
section the constrained Robertson transformation between ARF and an
ordinary IRF, which determines the covering theory.

We suppose there are two frames. One is isotropic ARF,
$\Sigma_0$, in which the one-way VL is constant, and the other is an
ordinary IRF, $\Sigma$. In the case that the observer is in
$\Sigma_0$, he/she lets the frame $\Sigma$ move with velocity $v_0$
in the x direction with respect to $\Sigma_0$. Meanwhile, in the
case that the observer is in $\Sigma$, he/she finds that the frame
$\Sigma_0$ moves with velocity $-v$ in the x direction with respect
to $\Sigma$. For the lack of symmetry, the result $v_0= v$ is not
held generally. However, we are entitled to assume that the x-axes
of the two frame are parallel to each other at all time.

To any system of values $(x_0,y_0,z_0,t_0)$, which completely
defines space and time of an event in ARF, there is a system of
values $(x,y,z,t)$ determining that event in
 IRF $\Sigma$. Letting the axes of X in the two systems coincide,
and their axes of Y and Z be parallel respectively, we now want to
find the transformations connecting these two systems of values
which depict the same event.

After suitable choices of origins of space and time, the
transformation between the two frames has the following form
 \bee \label{trans}
 \begin{array}{l}
   x_0=a_{11}x+a_{14}t, \\
   y_0= g_1 y, \\  z_0=g_2 z, \\
   t_0=a_{41}x+a_{44}t.
 \end{array}
 \ee
Since the velocity of $\Sigma_0$ with respect to $\Sigma$ is $-v$,
we have $a_{14}=a_{11}v$ immediately. It is apparent that the
$\Sigma$ has a rotation symmetry with respect to x-axis, therefore,
$g_1=g_2$. Furthermore, VL in $\Sigma$ should have the form of
$c(\varphi)=\frac{1}{1+A\cos\varphi}$ with $\varphi$ the angle
between propagation direction and x-axis direction.

At time $t=t_0=0$, when the origins of the two frames coincide,
let a light flash be emitted therefrom, and be propagated with the
velocity unit in $\Sigma_0$. If $(x_0, y_0, z_0,t_0)$ is an
wavefront event in x-y plane, we have
 \bee \label{frame}
t_0=\sqrt{x_0^2+y_0^2}.
 \ee
 In frame $\Sigma$ such event is depicted by
 \bee \label{wavefont}
\frac{t}{1+A u}=\sqrt{x^2+y^2}=\frac{x}{u}, \ee
 with $u=\cos\varphi=\frac{x}{\sqrt{x^2+y^2}}$.
In other words,
 \bee \label{wavefont_prime}
 \tag{\ref{wavefont}$'$} t=(A+\frac{1}{u})x.
 \ee
 Substituting Eqs. (\ref{wavefont_prime}) and (\ref{trans}) into Eq.
(\ref{frame}) and noticing that Eq. (\ref{frame}) holds for
arbitrary $u$, one has
 \beea
g_1(g_2) &=& a_{11} \sqrt{(1+A v)^2-v^2}, \nonumber \\
a_{44} &=&  a_{11}(1+ A v), \nonumber \\
a_{41} &=& a_{11}(v-A(1+A v)).
 \eea
Since $a_{11}$ is in fact a scale shift, one can simply set
$g_1=g_2=1$  and therefore $a_{11}=\frac{1}{\sqrt{(1+Av)^2-v^2}}$ in
the above equations.

Thus, the transformations between ordinary IRF and ARF are
 \bee \label{trans1}
 \begin{array}{l}
   x_0=\gamma(x+vt), \\
   y_0= y, \\ z_0= z, \\
   t_0=\gamma[(v-A(1+A v))x+(1+A v)t],
 \end{array}
 \ee
 and
\bee \label{trans2}
 \begin{array}{l}
   x=\gamma((1+A v)x_0-vt_0), \\
   y= y_0, \\ z= z_0, \\
   t=\gamma[-(v-A(1+A v))x_0+t_0],
 \end{array}
 \ee
with $\gamma=\frac{1}{\sqrt{(1+Av)^2-v^2}}$. One can verify that in
$\Sigma_0$, the observer finds that the velocity of $\Sigma$ is
$v_0=\frac{v}{1+A v}$. The fact that $v_0 \neq v$  reflects the
asymmetry of the two frames. In the above equations A is the
function of $v_0$ (or $v$). Since $v_0=0$ or $v=0$ implies that the
two frames $\Sigma_0$ and $\Sigma$ should be the same, we conclude
that $A(v)=0$ at $v=0$ from Eqs. (\ref{trans1}) and (\ref{trans2}).
We therefore rewrite the transformations utilizing
$v_0$ and $A=Bv_0$ in the following
 \bee \label{trans1:prime} \tag{\ref{trans1}$^\prime$}
 \begin{array}{l}
   x_0=\gamma_0((1-B v_0^2)x+v_0 t), \\
   y_0= y, \\ z_0= z, \\
   t_0=\gamma_0((1-B)v_0x+t), \end{array}
 \ee
 and
\bee \label{trans2:prime} \tag{\ref{trans2}$^\prime$}
 \begin{array}{l}
   x=\gamma_0(x_0-v_0t_0), \\
   y= y_0, \\ z= z_0, \\
   t=\gamma_0[-(1-B)v_0x_0+(1-B v_0^2)t_0],
 \end{array}
 \ee
with $\gamma_0=\frac{1}{\sqrt{1-v_0^2}}$. Eqs.
(\ref{trans1})-(\ref{trans2:prime}) can be regarded as the
generalized Lorentz transformation. Therefore, the transformations
between velocity
 $(u_{x0},u_{y0},u_{z0})$ in $\Sigma_0$ frame and velocity $(u_{x},u_{y},u_{z})$
in $\Sigma$ frame are
 \bee \label{trans3}
 \begin{array}{l}
   u_{x0}=\frac{(1-Bv_0^2)u_x+v_0}{(1-B)v_0u_x+1},  \\
   u_{y0}=\frac{\gamma_0^{-1} u_y}{(1-B)v_0u_x+1},  \\
   u_{z0}=\frac{\gamma_0^{-1} u_z}{(1-B)v_0u_x+1}, \\
  \end{array}
 \ee
 and
\bee \label{trans4}
 \begin{array}{l}
   u_x=\frac{u_{x0}-v_0}{-(1-B)v_0u_{x0}+(1-Bv_0^2)}, \\
   u_y= \frac{\gamma_0^{-1} u_{y0}}{-(1-B)v_0u_{x0}+(1-Bv_0^2)}, \\
   u_z= \frac{\gamma_0^{-1} u_{z0}}{-(1-B)v_0u_{x0}+(1-Bv_0^2)},
 \end{array}
 \ee
respectively.

We have thus shown the most generally possible form if we only take
the assumption of close-path VL invariant. Parameter $B$ depends on the
velocity $v$ (or $v_0$) and the different dependency determines
different theory.

From Eqs. (\ref{trans1:prime}) and (\ref{trans2:prime}), both the
Lorentz contraction effect and the time dilation are valid if and
only if the observer stays in ARF, $\Sigma_0$. These results,
although from different starting-point, are in agreement with those in
refs. \cite{selleri,sexl,russo,serr2}. It is obvious that our
deductions are very concise.   Furthermore, although theories are
possibly very different, $\gamma_0=\frac{1}{\sqrt{1-v_0^2}}$ means
there is no superluminal object in ARF. Such statement is apparently
valid in other IRF. Therefore, the closed-path VL
invariant implies there is no superluminal signal. The
communication with the past is then impossible, {\it i.e.} there is no
Tolman's paradox\ct{tolman}. Noticing that here we make only the
assumption of the closed-path VL invariant, we think the result is very
significant.

For all the equivalent theories, we have two conventional choices:
\begin{enumerate}
  \item One may choose $B=0$. In this theory the one-way VL is constant in
  all the inertial reference frames. In other words, all the IRF
  are equivalent and we return to the special theory of
  relativity.
  \item One may also choose $B=1$\ct{serr2}. Now we get $t_0=\gamma_0 t$. That
  is, there is a universal time in all the IRFs, including ARF. In
  this theory the simultaneity is absolute although there is time dilation.
  However, VL is anisotropic in an ordinary IRF except ARF. One may think that all the
  cosmological phenomena, such as, extragalactic galaxies redshift, microwave
  background radiation, are respect to ARF.
\end{enumerate}

Therefore, the relativity postulate and the absoluteness of
simultaneity can not be valid in one theory. These two concepts can
only be valid separately, that is, they can be valid in different
theories. The only difference between these two theories is the
different choice of the function $B$. Due to the fact that such
choice makes no physical observation, we conclude that these two
concepts are not inconsistent\ct{selleri,russo}.

\section{Conclusions}
Taking closed-path VL invariant as the starting-point,
we discuss here the general form of VL. We find that such postulate
puts a very strong constraint on VL and that there is only one free
parameter in the expression of VL.

We also show the covering theory of special relativity under such
postulate without relativity postulate. Since the sole parameter $B$
determines the covering theory, there are many similarities among
the covering theories. The two significant similarities are: 1)
There is no superluminal signal; 2) the Lorentz contraction effect
and time dilation are valid partially except the special relativity,
that is, these two concepts are valid in other theories only if the
observer stays in ARF, $\Sigma_0$. However, the concept of
relativity of simultaneity in the special theory of relativity is
not universal. In other words, different covering theory has
different opinion on the relativity of simultaneity. In particular,
we have shown in the paper a special theory in which the
simultaneity is absolute. Indeed, our study shows that the two
concepts, relativity postulate (one-way VL invariant) and the
absoluteness of simultaneity, can not be satisfied simultaneously in
one theory. One can freely adopt conventional choice on different
problem. But since such choice takes no physical effect, these two
concepts are not inconsistent.

This work is supported by the Cultivation Fund of the
Key Scientific and Technical Innovation Project-Ministry
of Education of China (No. 708082) the Foundation of Xi'an Jiaotong University.

\vskip 0.2in

%\balance

\end{multicols}


\begin{thebibliography}{99}
\bibitem{einstein}
A. Einstein, {\it Ann. d. Physik} {\bf 18}, 639 (1905).

\bibitem{stachel}
J. Stachel, {\it Nature} {\bf 433}, 215 (2005).

\bibitem{holton}
G. Holton, {\it Nature} {\bf 433}, 195 (2005).

\bibitem{franklin}
J. Franklin, {\it Eur. J. Phys.} {\bf 31}, 291 (2010).

\bibitem{bondi}
H. Bondi, {\it Observatory(London)} {\bf 82}, 133 (1962).


\bibitem{bergman}
P. Bergmann, {\it Found. Phys.} {\bf 1}, 17 (1970).

\bibitem{5}
K. Scharnhorst {\it Phys. Lett.} {\bf B236}, 354 (1990); G. Barton, {\it Phys. Lett.} {\bf B237} ,
559 (1990).

\bibitem{6}
 A. Enders and G. Nimtz, {\it J. Phys.}, {\bf I2}, 1693 (1992); G. Nimtz, {\it Eur. Phys. J.} {\bf B7},
523 (1999).

\bibitem{7}
L.J. Wang, A. Kuzmich and A. Dogarlu, {\it Nature}, {\bf 406}, 277 (2000).

\bibitem{8}
J.J. Carey {\it et al}, {\it Phys. Rev. Lett.} {\bf 84}
, 1431 (2000).

\bibitem{9}
D. Mugnai, A. Ranfagni and R. Ruggeri, {\it Phys. Rev. Lett.,} {\bf 84}, 4830 (2000).

\bibitem{rel}
Translation from The Collected Papers of Albert Einstein, vol. 2, The Swiss Years: Writings 1900-1909, English Translation (Princeton University Press 1989), p. 257

\bibitem{vel}
Translation from The Collected Papers of Albert Einstein, vol. 2, The Swiss Years: Writings 1900-1909, English Translation (Princeton University Press 1989), p. 124

\bibitem{levy}
J. Levy, {\it Found. Phys.} {\bf 34}, 1904 (2004).

\bibitem{covering}
F. Tangherlini, {\it Nuovo Cimento Suppl.}{\bf 20}, 1 (1961).

\bibitem{sexl}
R.M. Mansouri and R.U. Sexl,{\it Gen. Rel. Grav.} {\bf8} 497; 515;
809  (1977).

\bibitem{selleri}
F. Selleri, {\it Found. Phys.} {\bf 27}, 1527 (1997).


\bibitem{russo}
G. Russo, {\it IL Nuovo Cimento,} {\bf B121}, 65  (2006).


\bibitem{serr2}
F. Selleri, {\it Found. Phys.} {\bf 26}, 641 (1991); {\it Found. Phys. Lett.} {\bf 9}, 43 (1996).

\bibitem{vargas}
J.G. Vargas and D.G. Torr, {\it Found. Phys.} {\bf 16}, 1088 (1986).

\bibitem{tolman}
R.C. Tolman, {\it The Theory of the Relativity of Motion} (University of California Press,
Berkeley, 1917).

\end{thebibliography}
\end{document}